# Use of survival analysis and simulation to improve maintenance planning of high voltage instrument transformers in the Dutch transmission system

Swasti R. Khuntia[1], Fatma Zghal[1], Ranjan Bhuyan[1], Erik Schenkel[1], Paul Duvivier[2], Olivier Blancke[2], Witold Krasny[2]

**Abstract** This paper describes the use of survival analysis and simulation to model the lifetime of high voltage instrument transformers in the Dutch transmission system. To represent asset aging, the non-parametric Kaplan-Meier method is used to enable the fitting of Weibull distribution. Such an approach is implemented on three different voltage levels, namely 110kV, 150kV, and 220/380kV. Real failure and inspection data is used to achieve a realistic failure model of the instrument transformers. Failure and maintenance data occurring between 1989 and 2021 have been used for this study. In spite of missing and low-quality data, a rich failure database could still be prepared. This study also offers insights into factors (i.e., voltage level, in-service age) influencing the remaining life from both graphical survival function and parametric Weibull distribution analysis. Based on the derived statistics, future possible maintenance planning scenarios are simulated under a complex system modelling framework in a digital twin enabled platform. Eventually, the scenarios are evaluated in terms of replacement costs (CAPEX), inspection hours, and unavailability hours.

## 1 Introduction

TenneT, as European transmission system operator, is facing power supply reliability challenges that originate in a globally aging infrastructure and increasing complexity of business operations in the context of energy transition. While power transformers, due to the criticality of their function on the grid have been the focus of many studies, concerns have been raised recently on the lack of focus on long-term asset management of Instrument Transformers (ITs). ITs play an important role in the metering of electrical quantities and protection of other system components. Due to their importance, any unplanned unavailability due to failures can cause considerable outage costs to utilities. Consequently, it is crucial to properly characterize the aging of ITs using statistical approaches that will enable to predict the evolution of the IT population failure over the next years. In addition, it will yield valuable perspectives in terms of optimizing maintenance and replacement policies accordingly. The reliability analysis of ITs is very much dependent on the defined maintenance strategies which will provide a reliable and safe power supply. By definition, asset management involves strategies to explore, identify, plan, invest, utilize, maintain, replace, and dispose of assets while maximizing their value and performance under some prescribed financial constraint (Khuntia et al., 2016). Since ITs play such an important role, it is expected that statistical failure analysis will give a better insight on actual maintenance planning performance to the asset management team at TenneT. Technically, in the reliability analysis of IT, it is interesting to identify the independence or dependence of the specific covariates that indicate the operation of the IT.

[1] S.R. Khuntia (✉), F. Zghal, R. Bhuyan, E. Schenkel
Asset Management Onshore, TenneT TSO B.V., Arnhem, The Netherlands
e-mail: firstname.lastname@tennet.eu

[2] P. Duvivier, O. Blancke, W. Krasny
Cosmo Tech, Lyon, France
email: firstname.lastname@cosmotech.com

For any kind of data-driven methodology and, in particular, asset reliability characterization, a robust database is needed, both in terms of volumetry and quality (Balzer and Neumann, 2011). However, it can be argued that there should be a preference for robust data and that there are techniques that could be used to cope with data discrepancies. In our case, the historical failure data play an important role in understanding the behavior of ITs. Literature study reveals that explosion is one of the highest reported failure modes. Impact of explosion not only relates to direct cost of IT replacement but also chances of replacement of neighboring equipment damaged in the explosion. CIGRE reports are one of the primary sources for publicly available failure databases of ITs. Three series of CIGRE reports are available online. The first report was published in 1990 which covered failures of ITs (voltage >72.5kV) in about 15 countries. The survey covered 136033 transformers in the period from 1970 to 1986 (CIGRE, 1990). The second report published results for 131207 ITs (voltage > 60kV) in the period from 1985 to 1995 in the year 2009 (CIGRE, 2009). The third results of a wider international survey was published in 2012. It collected population and failure data for ITs of voltage > 60kV and excluded AIS ring current transformers that were in service during the years 2004 to 2007 inclusive (CIGRE, 2012). Some other failure investigations were reported (Poljak et al., 2010; Raetze et al., 2012; Tee et al., 2021), where authors focus on reduction of IT explosion and better condition monitoring of ITs. Nonetheless, the truth is that failure is probabilistic in nature, and it needs investigations on the relationship with asset data and failure cause. The use of semi-parametric Cox model was reported in (Tee et al., 2021). The authors elaborated the factors influencing the probability of failures through analysis on the lifetime data from both graphical survival function plots and semi-parametric Cox model.

With the use of Simulation Digital Twin technology from Cosmo Tech, TenneT analyzed various maintenance strategies. The Digital Twin has been calibrated based on the historical failure data that it recorded with statistical technique relying on survival analysis. Literature study shows that survival analysis was used for power transformer reliability studies of around 2000 nos. in the Canadian and around 6000 nos. in the Australian utility (Picher et al., 2014; Martin et al., 2018). Ref. (Picher et al., 2014) described the data of Canadian utility Hydro-Quebec where they adopted a good match using the Kaplan-Meier and Weibull distribution. Finally, the method concluded that Weibull distribution is a better fit and the results looked promising. Similarly, ref. (Martin et al., 2018) followed a similar strategy for Australian data. The authors deduced the choice of Kaplan-Meier or Weibull distribution based on the different voltage classes. In practice, Weibull distribution fitted to empirical failure data are commonly used to calculate life expectancy. However, the challenge in applying such a distribution to electrical assets is that often the root cause of failure is not related to the normal aging of the asset, but rather external factors. The aim of this paper is three-fold: (1) use of real failure data to model a time-varying failure rate based on Weibull parameters obtained from Kaplan-Meier survival analysis, (2) investigate extrapolation methods to maximize value of existing inspection results across IT population, and (3) use digital twin enabled simulation to tune the required resources necessary to realize TenneT's strategy for considered substation equipment maintenance and renewals.

## 2   Data and Methodology

### 2.1   Description of Data

As of the date of writing this paper, TenneT owns and maintains a large fleet of ITs in the Dutch high voltage AC network (i.e., 110, 150, 220 and 380kV) as shown in Figure 1(a). It is of interest to see the age profile of the existing population, in terms of years since manufacture because reliability is often related to age. However, lifetime data can be complicated as some ITs often extend over several decades. At TenneT, the expected design life of an IT is 45 years. This age is affected and reduced, sometimes substantially, depending on the design or utilization of the IT, i.e. its loading or the environment to which it is exposed. In some cases, a good maintenance scheme can even increase the replacement age. Although there is no prescribed replacement age, it is the responsibility of the asset management department to formulate the maintenance policies based on failure history. For this study, failure data was obtained from various sources, starting from failure records, reports to talking to experts. Fortunately, TenneT did not record a high number of major failures since the 1989. A major failure is defined as a sudden explosive event that has caused an immediate emergency system outage or trip. Figure 1(b) lists the failure events with respect to manufacturer (coded for confidentiality) and IT age.

The failure list was not adequate to come up with a statistical model. In addition, maintenance reports (or work orders) and expert knowledge was used to populate the list and gain utmost information. A work order is a document that provides all S.R.Khuntia - Use of survival analysis and simulation to improve maintenance planning of high voltage instrument transformers in the Dutch transmission system    4 the information about a maintenance task and outlines a process for completing that task. In case of IT, corrective work orders are used (the others being periodic maintenance and inspection work orders). Discussion with experts led us to use the work orders when an IT was out of service for any kind of maintenance. Figure 1(c) shows the total recorded failures for the IT population. In the recent years, one observation worth noticing is that the number of failures has increased significantly.

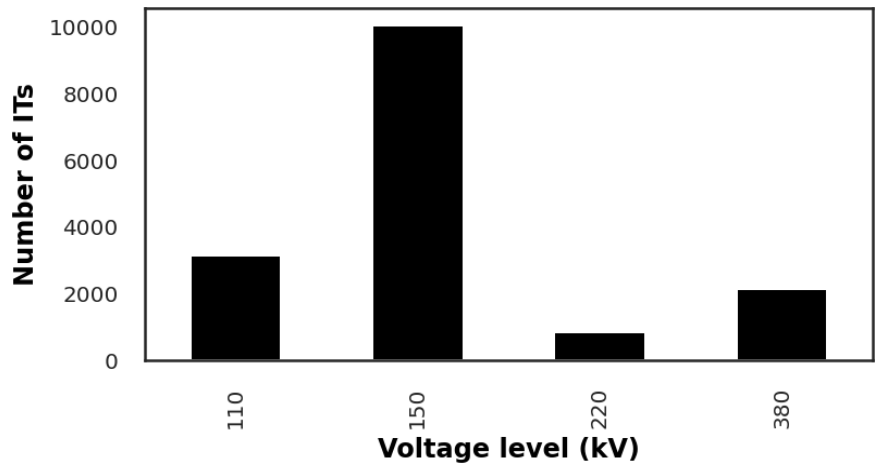

*(a)*

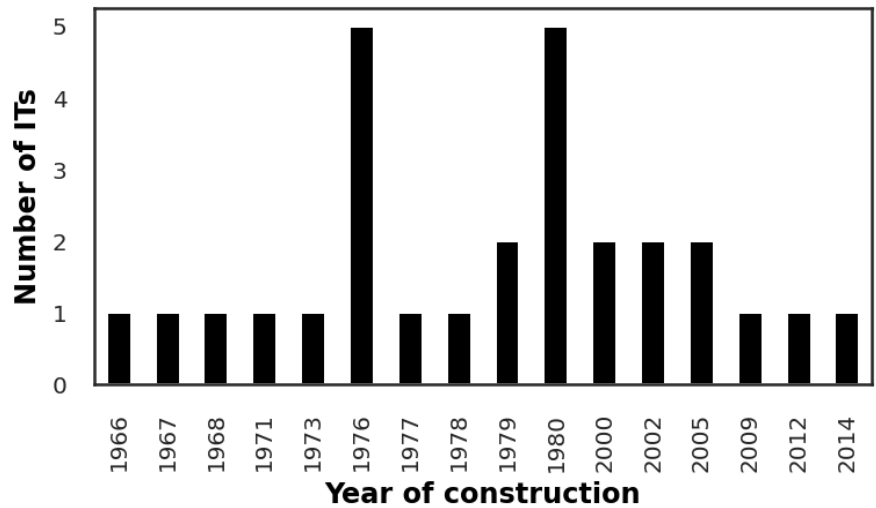

*(b)*

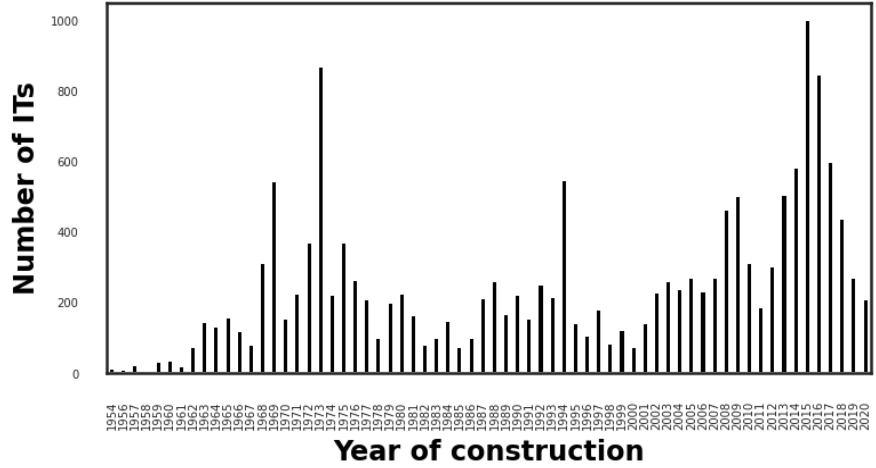

*(c)*
*Figure 1 (a) Voltage-based IT population, and (b) Actual failure list until July 2021, (c) Populated failure from work order and expert opinion until July 2021*

## 2.2 Survival Analysis and Failure Rate Modelling

Survival analysis is a statistical technique used to estimate the lifespan of a particular population under study. It is an analysis of time-to-event data (Wang et al., 2019). One of the widely used survival analysis technique is the Kaplan-Meier (KM) estimate (Bland and Altman, 1998). The KM estimator uses lifetime data to perform survival analysis. Although it is widely used in medical research to gauge the part of patients living for a specific measure of time after treatment, it has been used in the power systems sector to model the survival of electric assets (Martin et al., 2018). The use of KM estimate is supported by two reasons: one is that it does not assume that the data fits a statistical distribution, and second is that it allows the inclusion of censored data (when an IT had not failed by mid-2021).

For a population, the survival function $\hat{S}(t)$ is defined as:

$$\hat{S}(t) = \prod_{i:t_i<t} \left(1 - \frac{d_i}{n_i}\right)$$

where, $t_i$ is the time at least one event happened, $d_i$ is the number of events that happened at time $t_i$ and $n_i$ is the number of individuals known to have survived up to time $t_i$ (Davidson-Pilon, 2019). In our study, the estimates are calculated for three different voltage levels and $n_j$ considers observations that occurred between the oldest IT age and mid-2021. An important aspect in survival analysis is considering the censored data. Censoring occurs when the value of an observation is only known to some extent. Censored data is often encountered when analysing practical life data, especially in case of electrical power systems where most of the installed equipment is still in-service, and most of the time the exact age of equipment at the moment of failure is unknown (CIGRE, 2017). In this study, a large amount of data falls under the right censored data (suspended data) category. A dataset is termed as right censored or suspended when it is composed of components that did not fail. The term right censored indicates that the event is located to the right of the dataset, which implies that certain components are still operating. In our dataset, we had to deal with right censoring and no left truncation since the year of construction was known to us. Ignoring truncation causes bias in model's estimation.

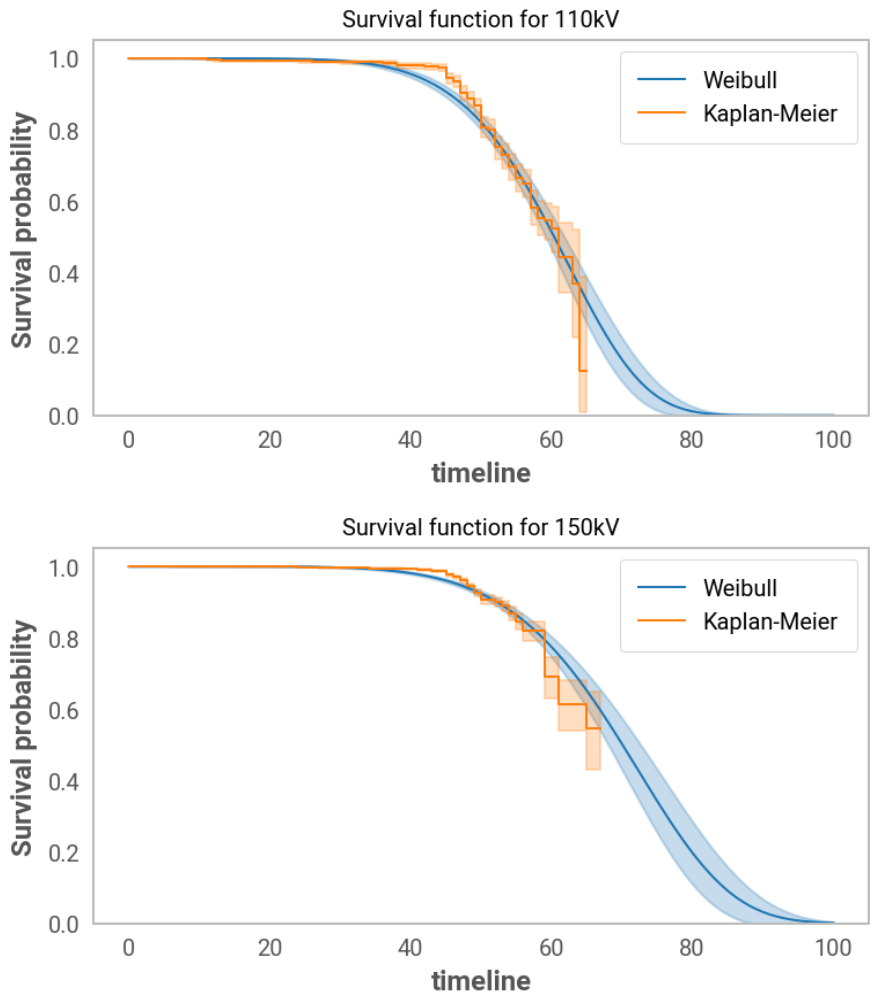

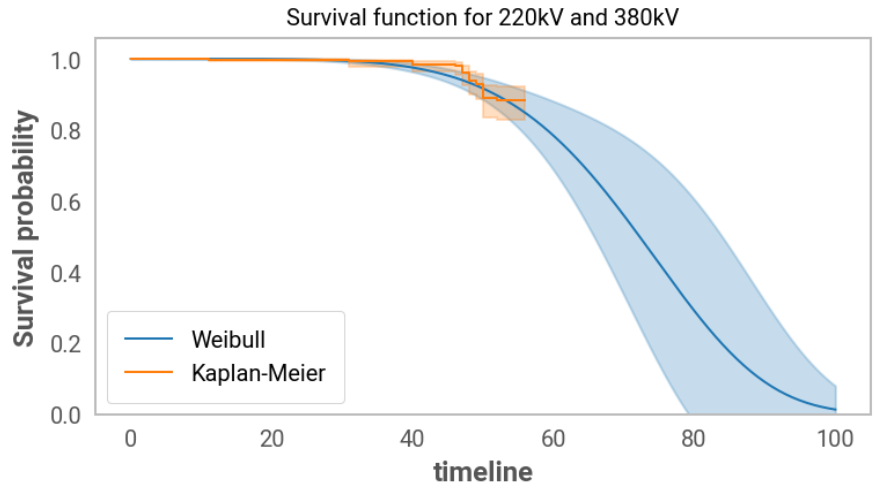

*Figure 2 Kaplan-Meier estimate of all different voltage levels.*

The IT dataset was split into three different families, each one with its own degradation law, based on their voltage level as is shown in Figure 2. A useful statistic in this analysis is calculating the median survival time, which defines the point in time where on average 50% of the population should have failed. For 110kV, the median survival time is 61 years. However, the median survival time for 150, 220 and 380kV is infinity because there have been an insufficient number of failures to determine it. In such cases, the two best options are:
1. use another quantile (e.g. 0.75) to compare the groups;
2. approximate the survival curve by means of a parametric fit and derive the median survival time using the model.

The second option is chosen in our study since all the three voltages can be modelled using the parametric fit assuming that failure times have a Weibull distribution. In other words, Weibull distribution is used to parameterize the KM estimate. The Weibull distribution is a widely used method to analyse the statistical features of failure (Rinne, 2008). The probability $f(t)$ and cumulative density function $F(t)$ are defined as: $f(t) = \beta \frac{t^{\beta-1}}{\eta^\beta} e^{-\left(\frac{t}{\eta}\right)^\beta}$ and $F(t) = 1 - e^{-\left(\frac{t}{\eta}\right)^\beta}$ ; where, $t$ is the time, $\beta$ is the shape and $\eta$ is the scale parameter. Table 1 shows the different parameters calculated for our study from the corresponding survival function.

*Table 1 Statistics and Weibull parameters.*

| Voltage (kV) | No. of ITs | No. of censored | β | η | median |
|---|---|---|---|---|---|
| 110 | 3168 | 255 | 6.67 | 63.79 | 61 |
| 150 | 10058 | 298 | 6.42 | 74.20 | infinity |
| 220 and 380 | 2982 | 25 | 5.65 | 77.05 | infinity |

## 3    Modelling in Cosmo Tech Asset and Simulations

Founded in 2010, Cosmo Tech is a technology company pioneer in the modeling of complex systems (https://cosmotech.com/). Relying on its industry-validated modeling and simulation software platform, Cosmo Tech has developed a solution called Cosmo Tech Asset, henceforth called CTA. CTA allows to build digital twins of asset portfolios with their full complexity such as network dependencies, operative strategies, or dynamical resources allocations.

### 3.1    Cosmo Tech Asset Platform

The different steps involved in the CTA platform are:
1. Experiment the CTA platform's pre-built health assessment methods and compare the results with internal initiatives. For health assessment, the asset health index is a key simulation variable, and it is described in the next sub-section.
2. Demonstrate the calibration of reliability law (using Weibull distribution) for simulations against up-to-date condition of ITs, but also historical IT related data, such as field observation or inspection data and measurement inputs.
3. Investigate the functional possibilities that would allow to leverage existing inspection results across ITs using extrapolation methods when applicable, therefore maximize inspection result value.
4. Finally, based on the achieved health assessment technique, use the simulation platform to tune the required resources necessary to realize TenneT's strategy for considered IT maintenance and replacements.

### 3.2    TenneT Asset Health Index

For health assessment, the TenneT asset health index (AHI) is considered and is shown in Table 1(a) (TenneT, 2021). The AHI is based on asset age and failure probability, and it is used to drive short-term maintenance and long-term replacement strategies. It provides a consistent way to compare the overall asset health of TenneT's assets.

The evaluation of the AHI is based on two metrics:
1. *probability of failure* of IT in the coming years for AHI score of 1 to 6, and
2. *age* of IT for AHI score of 7 to 10.

In addition to AHI, the study of IT uses reliability law over which failures are drawn during the simulations. The reliability law corresponds to the KM survival function and the Weibull estimates that are described in section 2. These laws have a cumulative distribution function which represent the probability for a failure to occur before a certain age. And the probability of failure over the next year can be evaluated using the following formula:

S.R.Khuntia - Use of survival analysis and simulation to improve maintenance planning of high voltage instrument transformers in the Dutch transmission system                                9

$$P(X < t+3 \mid X > t) = 1 - P(X > t+3 \mid X > t)$$
$$= 1 - \frac{P(X>t+3 \cap X>t)}{P(X>t)} = 1 - \frac{P(X>t+3)}{P(X>t)}$$
$$= 1 - \frac{1 - P(X<t+3)}{1 - P(X<t)} = 1 - \frac{1 - F(t+3)}{1 - F(t)}$$

where,
- $F$ is the cumulative distribution function of the reliability law
- $X$ is a random variable representing the occurrence of a failure.

*Table 1 (a)TenneT Asset Health Index (AHI) definition, (b) Classification of Resources (FTE: Full Time Employment).*

| AHI Score | Colour | Definition |
|---|---|---|
| 1 | Purple | Within 3 years, 80% of chance that the asset is irreparably damaged |
| 2 | Purple | Within 3 years, 50% of chance that the asset is irreparably damaged |
| 3 | Purple | Within 3 years, 20% of chance that the asset is irreparably damaged |
| 4 | Red | Within 7 years, 80% of chance that the asset is irreparably damaged |
| 5 | Red | Within 7 years, 50% of chance that the asset is irreparably damaged |
| 6 | Red | Within 7 years, 20% of chance that the asset is irreparably damaged |
| 7 | Orange | Older than 75% of the average age |
| 8 | Orange | Between 60% and 75% of the average age |
| 9 | Green | Older than 5 years old and less than 60% of the average age |
| 10 | Green | Younger than 5 years old |

*(a)*

| Activity name | Duration (h) | Required FTE | Material cost (€) | Workforce cost (€) | Total cost (€) |
|---|---|---|---|---|---|
| Inspection every 3 years | 0.5 | 1 | 0 | 41.624 | 41.624 |
| Inspection every 6 years | 1.33 | 2 | 49.81 | 180.18 | 229.99 |
| Replacement IT 110kV | 40 | 10 | 8211 | 35000 | 43211 |
| Replacement IT 150kV | 40 | 10 | 10044 | 35000 | 45044 |
| Replacement IT 220kV | 40 | 10 | 15000 | 35000 | 50000 |
| Replacement IT 380kV | 40 | 10 | 15000 | 35000 | 50000 |

*(b)*

## 3.3 Simulation

The reliability law was used to evaluate the different scenarios for an efficient maintenance planning. A simulation period of 100 years is chosen for this study since it is assumed that the most recent IT replacements will be in operation until the end of this century. Time-based scenario is the current maintenance planning at TenneT. It is compared against a condition-based scenario. Both the scenarios are explained in detail in Table 2. The resources are listed in Table 1(b).

*Table 2 Different Scenarios under Study.*

|  |  | Condition-based | Time-based |
|---|---|---|---|
| Replacement | 220/380kV | 45 years | 45 years |
|  | 110/150kV | AHI score red or purple | 45 years |
| Inspections on bay every 3,6,12 months | 220/380kV | No inspections | No inspections |
|  | 110/150kV | Time-based starting at 25 years | Time-based starting at 25 years |

In principle, both scenarios are very similar in the sense that the same simulation model dataset is used. The difference lies in the trigger for the replacement activities of the 110/150kV assets. In fact, in time-based scenario, which represents the current way of working, the trigger is based on the real age of the asset. As soon as the asset reaches 45 years of age, replacement is triggered, and action is performed as resources are unlimited. On the other hand, in the condition-based scenario, the trigger is based on the apparent age of the asset. The apparent age is an attribute of every asset that reflects its degradation rate and it can be different from the real age of the asset. If the apparent age is higher than the real age, the asset degrades faster than normal. If the apparent age is lower than the real age, the asset degrades slower than normal. When the apparent age of the asset reaches 50 or 54, it means that the asset is reaching AHI score of respectively 6 or 3 that is red or purple (see Table 1(a)), and the replacement action is triggered.

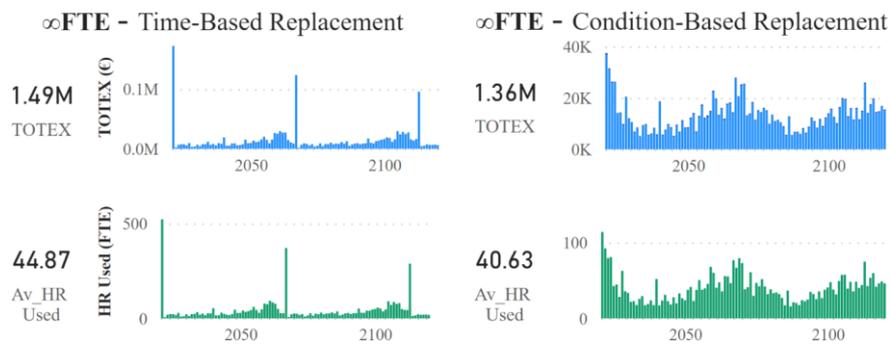

Figure 3 Unconstrained Scenarios Simulation.

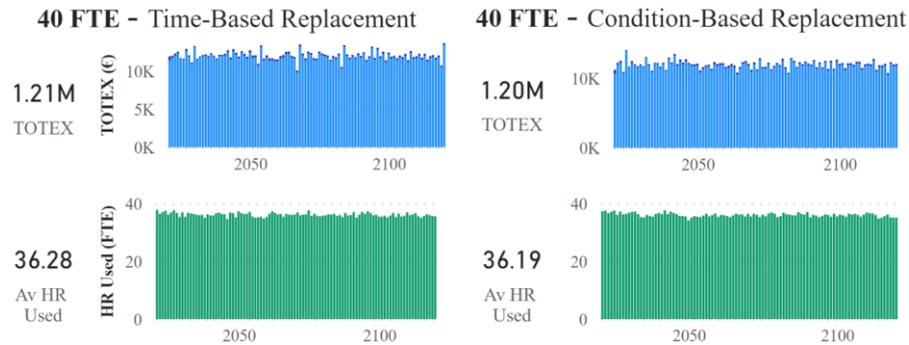

Figure 4 40 FTE constrained Scenarios Simulation.

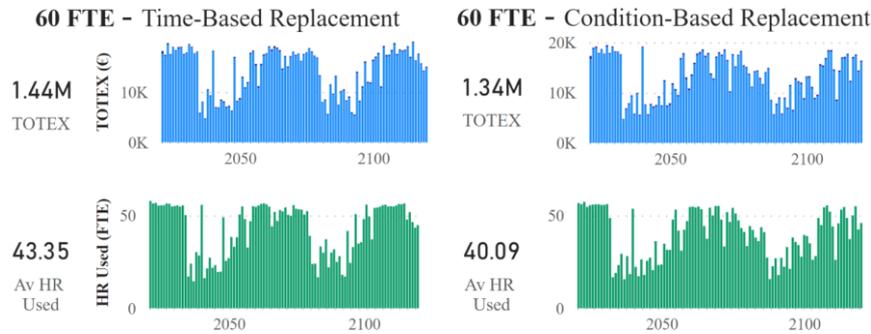

Figure 5 60 FTE constrained Scenarios Simulation.

From the figures, two conclusions can be made: (1) replacement activities are the major cost driver in the TOTEX (*Total Expenses*), and (2) Human resources (HR) costs are the major cost driver in the replacement costs. Simulation results show that in case HR availability is restricted, there is no significant difference between the time-based and condition-based replacement strategies. In fact, switching to a condition-based strategy might not be beneficial in that case since it comes with change and investments for little to no reward. If HR availability is guaranteed for the foreseeable future, then it is highly beneficial to switch from a time-based replacement strategy to a condition-based strategy as this would contribute to flattening the curve. Also, this would represent a lot of work at the beginning to prepare the necessary processes and investments for the new strategy but would lead to significant gains on the long term.

## 4   Conclusion

Maintenance planning of high voltage ITs using real data from the Dutch transmission system operator was illustrated in this study. The study aimed at understanding how digital twin enabled technology along with failure data can help TenneT to make better future maintenance strategies. The strategies aimed at easing financial decisions related to replacements (in terms of flattening the replacement curve) and unavailability of ITs in the network. Working on real data uncovered several challenges including missing data (both quantity and quality) and outliers. The non-parametric Kaplan-Meier survival analysis helped in parameter estimation of Weibull distribution. TenneT data could be translated to the data format to be used in the digital twin CTA tool, meaning that our data could be easily adapted to other software platforms. It is worth to mention that in this study, both data ownership as well as data confidence did not hinder the progress. Data confidence was built upon although multiple data sources had to be aligned together. TenneT partnered with Cosmo Tech to build the data ownership philosophy for successful digital twin implementation for maintenance planning.